\documentclass{WileyMSP-template}

\usepackage{newtxtext,newtxmath}
\usepackage{graphicx}
\usepackage{url}
\usepackage{xr}
\usepackage{ragged2e}
\justifying
\begin{document}

\pagestyle{fancy}
\rhead{}

\title{Direct Imaging of Hydrogen-Driven Dislocation and Strain Field Evolution in a Stainless Steel Grain}

\maketitle


\author{David Yang$^*$}
\author{Mujan Seif}
\author{Guanze He}
\author{Kay Song}
\author{Adrien Morez}
\author{Benjamin de Jager}
\author{Dmytro Nykypanchuk}
\author{Ross J. Harder}
\author{Wonsuk Cha}
\author{Edmund Tarleton}
\author{Ian K. Robinson}
\author{Felix Hofmann$^*$}


\begin{affiliations}
Dr. D. Yang$^*$, Dr. M. Seif, Dr. G. He$^{\dagger}$, Dr. K. Song$^{\S}$, Mr. A. Morez, Dr. B. de Jager, Prof. E. Tarleton, Prof. F. Hofmann$^*$\\
Department of Engineering Science, University of Oxford, Oxford OX1 3PJ, UK\\
Email Address: david.yang@eng.ox.ac.uk, felix.hofmann@eng.ox.ac.uk\\

Dr. D. Yang$^*$, Prof. I. K. Robinson\\
Condensed Matter Physics and Materials Science Department, Brookhaven National Laboratory, Upton, NY 11973, USA\\

Dr. D. Nykypanchuk\\
Center for Functional Nanomaterials, Brookhaven National Laboratory, Upton, NY 11973, USA\\

Dr. R. J. Harder, Dr. W. Cha\\
Advanced Photon Source, Argonne National Laboratory, Lemont, IL 60439, USA\\

Prof. I. K. Robinson\\
London Centre for Nanotechnology, University College London, London WC1E 6BT, UK\\

$\dagger$Present address: Shanghai Nuclear Engineering Research and Design Institute, Shanghai 200233, China.\\

$^\S$Present address: School of Aerospace, Mechanical and Mechatronic Engineering, The University of Sydney, Sydney, NSW 2006, Australia.

\end{affiliations}


\keywords{Hydrogen embrittlement, dislocations, strain field, stainless steel, Bragg coherent X-ray diffraction imaging}

\begin{abstract}
Hydrogen embrittlement (HE) poses a significant challenge to the durability of materials used in hydrogen production and utilization. Disentangling the competing nanoscale mechanisms driving HE often relies on simulations and electron-transparent sample techniques, limiting experimental insights into hydrogen-induced dislocation behavior in bulk materials. This study employs in situ Bragg coherent X-ray diffraction imaging to track three-dimensional (3D) dislocation and strain field evolution during hydrogen charging in a bulk grain of austenitic 316 stainless steel. Tracking a single dislocation reveals hydrogen-enhanced mobility and relaxation, consistent with dislocation dynamics simulations. Subsequent observations reveal dislocation unpinning and climb processes, likely driven by osmotic forces. Additionally, nanoscale strain analysis around the dislocation core directly measures hydrogen-induced elastic shielding. These findings experimentally validate theoretical predictions and offer mechanistic insights into hydrogen-driven dislocation behavior. The quantified nanoscale phenomena serve as critical inputs for multiscale modeling frameworks to predict bulk material responses and accelerate the development of HE-resistant alloys.
\end{abstract} 

\section{Introduction}

Hydrogen is an attractive energy carrier that can be produced through methods such as water electrolysis or photoelectrochemical technologies using renewable resources \cite{Dawood2020}. This green hydrogen can be used to decarbonize industries that rely heavily on fossil fuels such as aviation \cite{Tiwari2024}, heavy transportation \cite{Oliveira2021}, industrial production \cite{Griffiths2021}, and power generation \cite{Dawood2020}. As countries strive to reach their Net Zero 2050 sustainability targets, the global production of green hydrogen is forecast to rise annually, reaching 49 million tons per year by 2030 \cite{IEA2024}. This necessitates the design of hydrogen fuel systems to store, transport and distribute hydrogen. A long-standing challenge is hydrogen embrittlement (HE), which occurs when hydrogen atoms diffuse into metals, reducing their ductility and strength, leading to premature failure \cite{Johnson1875}. Austenitic stainless steel (SS), widely used due to its good corrosion resistance and strength, is not immune to this phenomenon. Understanding the mechanisms behind HE is essential to designing the next generation of HE-resistant materials. The study of hydrogen--SS interactions at the nanoscale can provide a mechanistic grounding for multiscale models that seek to predict behavior at larger length scales \cite{Castelluccio2018,Li2015b}.

HE has been extensively studied at the macroscale through mechanical testing, revealing how cracks initiate and propagate in hydrogen-exposed material \cite{Beachem1972,Shih1988,Oriani1974,Lynch1988} and can lead to in-service failure. To explain macroscopic behavior, advances in characterization techniques have enabled the exploration of HE at sub-micron length scales, which has been dominated by electron microscopy (EM) experiments and simulations \cite{Martin2019,Chen2024a,Dong2022}. This combination of experimental results and theory has led to multiple proposed degradation mechanisms responsible for HE. Here, we will briefly discuss the two most prominent mechanisms. In hydrogen-enhanced localized plasticity (HELP) \cite{Beachem1972,Shih1988,Robertson2015,Robertson2009}, hydrogen atoms in the metal lattice accumulate at and near the dislocation cores \cite{Chen2020a} and reduce the elastic stress field, known as hydrogen elastic shielding \cite{Birnbaum1994}. This makes it easier for dislocations to move, leading to a localization of deformation, which drives premature material failure. The second mechanism is hydrogen-enhanced decohesion (HEDE) \cite{Pfeil1926,Oriani1974}, which states that hydrogen atoms accumulate at locations of high triaxial stress, such as grain boundaries or crack tips. The presence of hydrogen reduces the cohesive forces between atoms, weakening the atomic bonds and causing intergranular fracture. The interplay between HELP and HEDE is multifaceted and possibly synergistic \cite{Djukic2019,Robertson2015,Koyama2014}, though there are still many different proposed theories about the mechanisms involved in HE \cite{Robertson2015,Lynch2012,Nagumo2004}.

Explaining these complex mechanisms requires insights at the nanoscale. Hydrogen can be trapped by vacancies, interstitial sites, solutes, and precipitates, as revealed by thermal desorption spectroscopy (TDS) studies \cite{Choo1982a}. Hydrogen has been reported to decrease the formation energy of vacancies, leading to an increased vacancy concentration, known as the superabundant vacancies (SAV) model \cite{Fukai2003}. Vacancies can trap multiple interstitial hydrogen atoms, revealed by positron annihilation spectroscopy (PAS) \cite{Myers1992} and density functional theory (DFT) calculations \cite{Nazarov2010}. Though less energetically favored, hydrogen also occupies octahedral interstitial sites in face-centered cubic (FCC) metals \cite{Fukai2005}. Similarly, hydrogen binds to interstitial and substitutional solutes \cite{Myers1992}, but this trapping is small compared to vacancy-solute defects \cite{Li2023}. Precipitates such as carbides are also strong hydrogen traps, revealed locally using cryogenic atom probe tomography \cite{Chen2020a}, and some have been shown to increase hydrogen embrittlement resistance by absorbing hydrogen \cite{Bhadeshia2016}. 

The ability to detect hydrogen has greatly enhanced our understanding of these mechanisms. Hydrogen microprint techniques \cite{Ovejero-García1985} and secondary ion mass spectroscopy \cite{Takai1995} can provide a static visualization of hydrogen distribution. Scanning Kelvin probe force microscopy \cite{Evers2013}, neutron tomography \cite{Griesche2014} and electron-stimulated desorption \cite{Miyauchi2020} have been valuable for studying time-resolved hydrogen movement in various microstructures. Despite the availability of these techniques, hydrogen is still generally considered difficult to probe due to its small size, high diffusion rate and low level of interaction with electron probes.



Consequently, there has been a push toward probing in situ responses to hydrogen diffusion to understand degradation mechanisms. Recently, dislocation movement caused by hydrogen desorption has been reported by Koyama et al. \cite{Koyama2020}, where they visualised micron-scale dislocation movements using electron channel contrast imaging (ECCI) on a Fe-Mn–based alloy. Huang et al. \cite{Huang2023} revealed bow-out motions of screw dislocations in $\alpha$-Fe pillars undergoing compression tests in a gaseous hydrogen environment inside a transmission electron microscope (TEM). This is an extension to the numerous TEM experiments done by the Robertson group to study hydrogen-induced dislocation movement in foil samples \cite{Robertson2001,Sofronis2002,Robertson1984,Shih1988,Wang2014}. However, EM techniques only probe relatively thin subsurface volumes of a material ($\lessapprox 100$ nm) \cite{Williams2009}, and the question of how hydrogen interacts with a single dislocation within the bulk in situ is still experimentally challenging, not well understood \cite{Myers1992,Martin2019}, and primarily addressed using simulations \cite{Dong2022}. Furthermore, hydrogen elastic shielding has only been inferred based on the observations of dislocation pileup \cite{Ferreira1998}, but the hydrogen-modified strain fields have not been directly measured experimentally. This information would greatly enhance our understanding of dislocation-hydrogen interactions to accelerate the development of HE-resistant alloys.





\begin{figure}[!ht]
    \centering
    \includegraphics[width=0.85\linewidth]{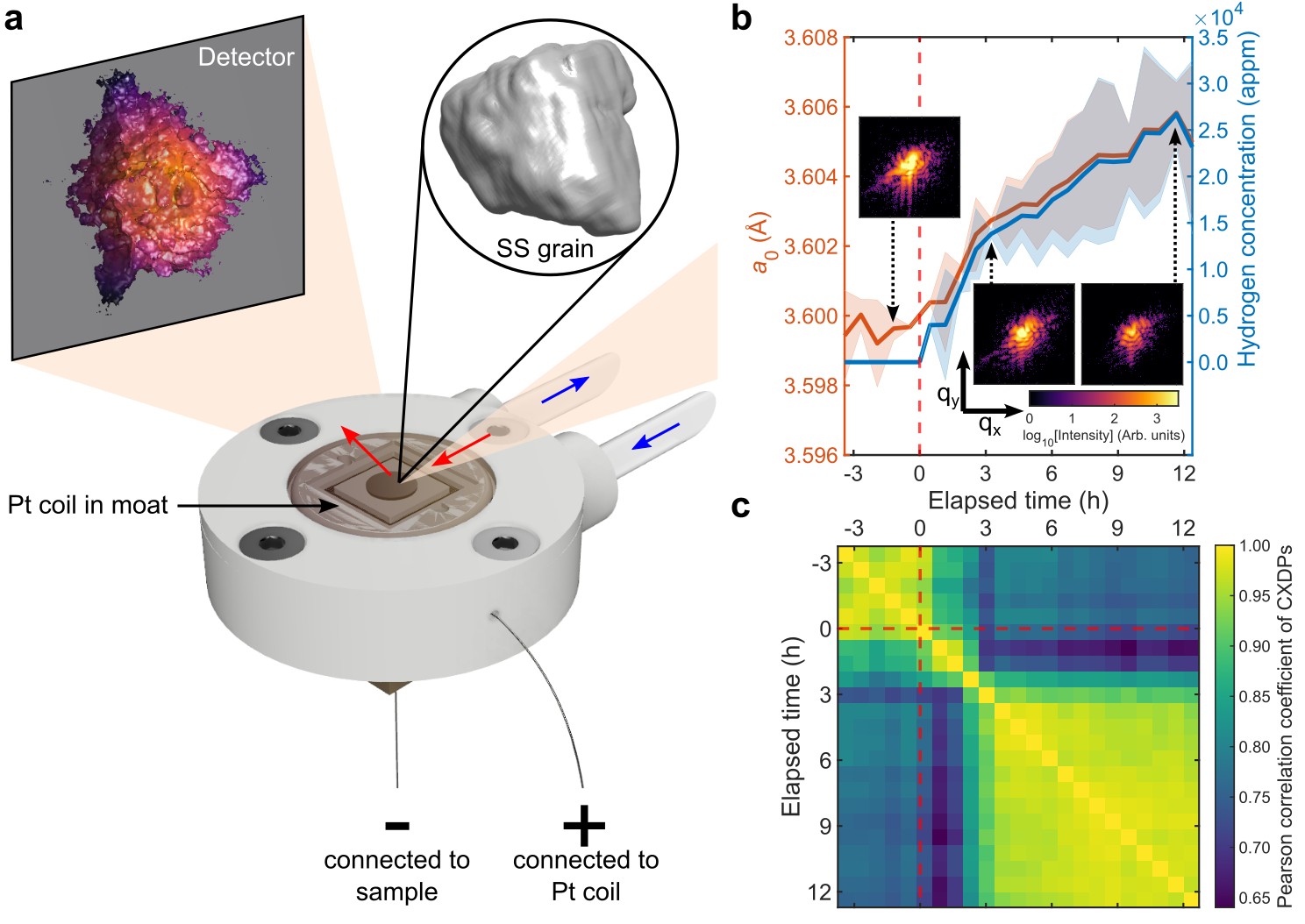}
    \caption{In situ hydrogen charging BCDI setup and evolution of the Bragg peak. a) The microcrystalline 316 SS disk is mounted in the electrochemical flow cell (see Supporting Information and Figure \ref{supp-fig:flowcell}, Supporting Information). An incoming coherent X-ray beam (red) illuminates a grain within the SS disk, and a slice through the reflected $111$ Bragg peak is captured on the detector. During in situ charging, a bias is applied to the SS (working electrode). A Pt wire coil (counter electrode) sits inside the flow cell moat. The blue arrows indicate the flow of the hydrogen charging solution. b) The lattice parameter and inferred hydrogen concentration before and during the experiment, with insets showing slices through the center of the Bragg peak corresponding to different times. The shaded areas correspond to the uncertainty associated with the Bragg peak position. c) 3D Pearson correlation of the Bragg peaks. Hydrogen charging starts at 0 h.}
    \label{fig:Fig1}
\end{figure}

Here we address these shortcomings by using in situ Bragg coherent X-ray diffraction imaging (BCDI) \cite{Pfeifer2006} to probe the evolution of a single dislocation and its associated strain field during hydrogen charging (\textbf{Figure \ref{fig:Fig1}}a), using a bespoke electrochemical flow cell (Supporting Information and Figure \ref{supp-fig:flowcell}, Supporting Information). BCDI is a nondestructive technique that captures three-dimensional (3D) nanoscale images of crystal defects and strain fields \cite{Clark2015,Orr2023,Singer2018,Yau2017b,Atlan2023a,Hofmann2020} by inverting coherent X-ray diffraction patterns \cite{Fienup1982} from a finite crystalline sample. To produce a sample with grains suitable for BCDI, we use high pressure torsion (HPT) \cite{Zhilyaev2008} to refine the grain structure of a 316 SS disk (composition shown in Table \ref{supp-table:composition}, Supporting Information). Powder X-ray diffraction data shows that this leads to a deformation-induced martensitic phase transition (Figure \ref{supp-fig:powder_XRD}, Supporting Information). The austenitic phase reappears after ex situ annealing, along with some controlled grain growth and defect relaxation (Experimental Section). The $111$ Bragg peak from an austentitic, FCC grain is measured before and during hydrogen charging at standard ambient temperature and pressure (Experimental Section). The center of the measured grain was close to the center of the disk and located $1.9\pm0.2\ \mathrm{\mu m}$ beneath the surface (Figure \ref{supp-fig:Grain_position}, Supporting Information). Phase retrieval algorithms are used to recover the real space electron density and phase from the diffraction data (Experimental Section), revealing information about the grain morphology, dislocations, and strain fields, $\varepsilon_{111}$, along the $[111]$ direction associated with the scattering vector, $\mathbf{Q}_{111}$. 

\section{Results and Discussion}

\subsection{Bragg peak evolution}\label{sec:Bragg peak evolution}

After initiating hydrogen charging at 0 h, we observe an increase in the average lattice parameter, $a_0$, (Figure \ref{fig:Fig1}b) determined using the center position of the Bragg peak (Experimental Section). There are several mechanisms that could explain this increase in $a_0$. It is favorable for hydrogen to be trapped in vacancies \cite{Myers1992}. However, the relaxation volume for a vacancy in FCC iron is negative and even with six hydrogen atoms per vacancy only becomes slightly positive \cite{Nazarov2010}. Thus, the accumulation of hydrogen-filled vacancies would not explain the substantial lattice swelling we observe. Alternatively, a Frenkel mechanism could be envisaged, where hydrogen prevents Frenkel defects from recombining, and the interstitial of the Frenkel pair (which has large positive relaxation volume) causes the observed swelling. However, this is unlikely because the formation energy of Frenkel defects is large. An energetically favorable explanation would be hydrogen occupying octahedral interstitial sites, which has been reported to cause lattice expansion \cite{Ulmer1993,Fukai2005}. Thus, we conclude that the homogeneous lattice strain observed, as a result of hydrogen charging, is dominated by a volumetric strain due to hydrogen occupying interstitial sites. Here, the homogeneous lattice strain, $e_{111}$, is relative to $a_0$ at --3.4 h before charging. Note that $e_{111}$ differs from $\varepsilon_{111}$: the latter is relative to the average lattice parameter at time $t$, $a_{0,\ t}$ (Experimental Section) \cite{Atlan2023a}. The volumetric strain is combined with the relaxation volume of hydrogen in austenitic SS to estimate the local hydrogen concentration (Experimental Section), which increases during charging (Figure \ref{fig:Fig1}b).

Figure \ref{fig:Fig1}b also shows central slices of the Bragg peak at different hydrogen concentrations as insets (see Figure \ref{supp-fig:Bragg peak slices}, Supporting Information for all central Bragg peak slices during charging). To quantify subtle changes, the 3D Pearson correlation coefficient (Experimental Section) is computed between the Bragg peak at different times, shown in Figure \ref{fig:Fig1}c. There is negligible change before 0 h, prior to hydrogen charging. Once hydrogen charging is applied at 0 h, the Bragg peak evolves until 3.2 h, after which it changes little. The evolution of the Bragg peak directly correlates to structural changes in the grain.

\subsection{Strain at the grain surface}\label{sec:Strain on the grain surface}
\begin{figure}[!ht]
    \centering
    \includegraphics[width=0.9\linewidth]{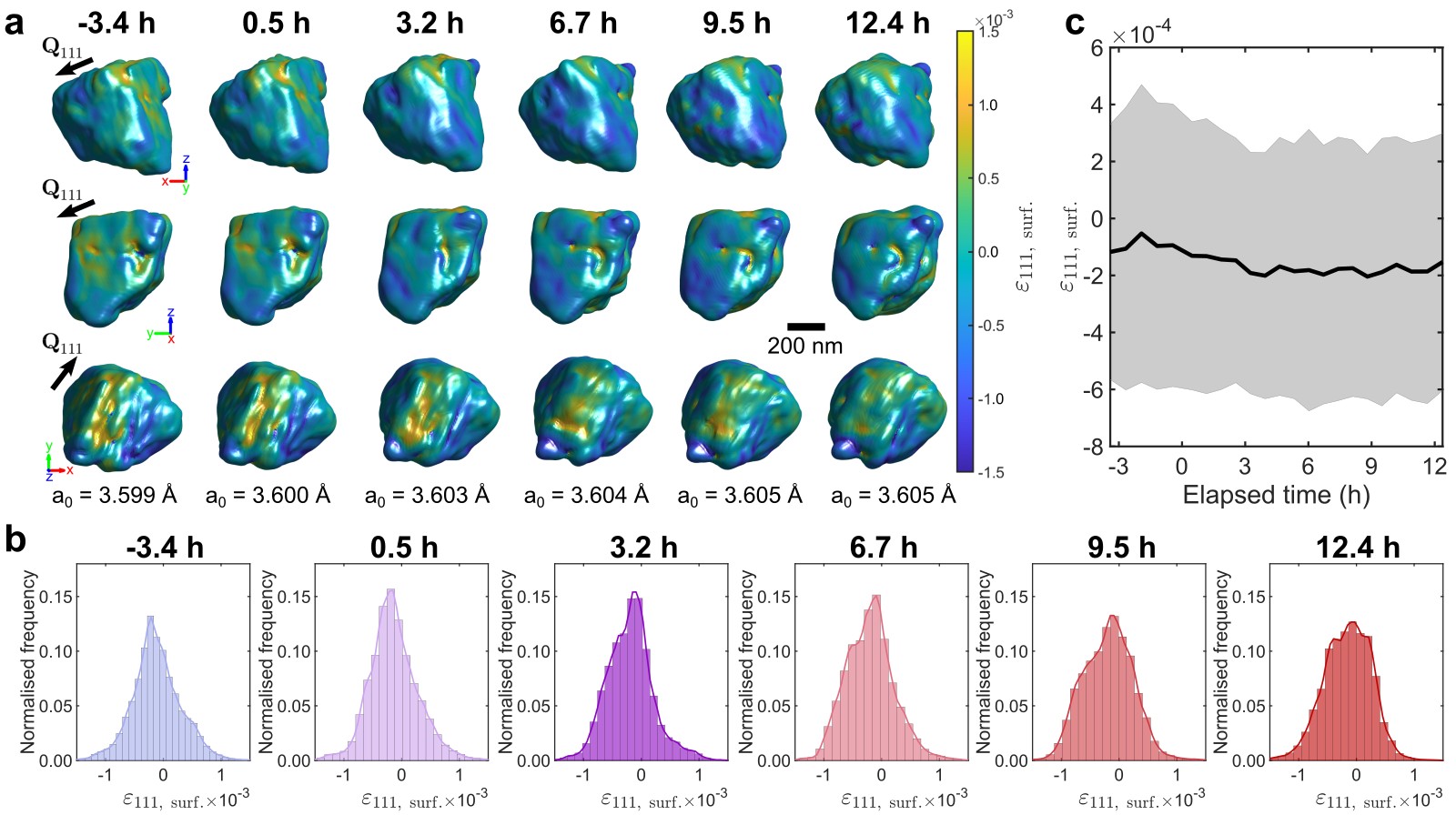}
    \caption{Evolution of the heterogeneous strain field on the grain surface before and during hydrogen charging. a) Different views of the grain surface, colored by $\varepsilon_{111,\ \mathrm{surf.}}$, the strain relative to the average lattice parameter indicated at the bottom. See Figure \ref{supp-fig:all isosurfaces}, Supporting Information and Video S1 for reconstructions from all time points. b) Histogram distribution of the surface strain for different time points during the hydrogen charging history. c) Evolution of average surface strain. The shaded region corresponds to one standard deviation.}
    \label{fig:Fig2}
\end{figure}

Figure \ref{fig:Fig1}b shows a large increase in average $a_0$ with hydrogen, leading to homogeneous strain. Does this homogeneous strain drive an increase in intergranular strain that could drive plasticity, or does hydrogen lead to a predominantly volumetric strain? If there were a substantial increase of intergranular strain, we would expect the surface strain of the grain of interest to increase due to mismatch with its neighbors. \textbf{Figure \ref{fig:Fig2}} shows the time evolution of the heterogeneous surface strain, relative to the average lattice parameter at each time. We refer to the surface strain, $\varepsilon_{111,\mathrm{surf.}}$, as the $\varepsilon_{111}$ value of the surface voxel of the reconstruction as determined using an amplitude threshold (Supporting Information). Note the surface strain is limited by our estimate of the 3D spatial resolution of 12 nm, determined using the phase retrieval transfer function (Supporting Information and Figure \ref{supp-fig:PRTF}, Supporting Information).

BCDI reconstructions are very sensitive to strain $\left(\approx 10^{-4}\right)$ and highly strained regions scatter outside the detector solid angle, so they appear as missing amplitude in the reconstructions \cite{Clark2015}. Relatively disordered grain boundaries with crystal defects in their vicinity appear rough in the reconstructions \cite{Yang2021}. This can be seen in Figure \ref{fig:Fig2}a, where the morphology is stable during the hydrogen charging, however, a softening of the grain edges is noticeable after 1.9 h (Figure \ref{supp-fig:all isosurfaces}, Supporting Information). Grain boundaries are strong hydrogen traps ($\approx 17\ \mathrm{kJ\ mol^{-1}}$)\cite{Choo1982a}, so they are expected accumulate hydrogen locally \cite{Kirchheim2007a} as recently shown by Chen et al. using cryogenic atom probe tomography \cite{Chen2020a}. As the grain boundaries becomes better organized, here facilitated by hydrogen, lattice distortion near the boundary reduces and hence the apparent surface reconstructed from BCDI evolves. We do not observe any grain boundary mobility due to the presence of new phases, as noticed in palladium thin film grains undergoing a phase transition under hydrogen partial pressures in a previous BCDI study \cite{Yau2017b}.


The distribution of $\varepsilon_{111,\ \mathrm{surf.}}$ in Figure \ref{fig:Fig2}a is shown in Figure \ref{fig:Fig2}b. The mean $\varepsilon_{111,\ \mathrm{surf.}}$ is slightly compressive at all times, and the distribution remains largely unchanged during hydrogen charging. This shows that hydrogen has little effect on grain-grain misfit and thus intragranular strains that could drive plasticity. Rather, hydrogen predominantly leads to a large volumetric strain compared to $\varepsilon_{111,\mathrm{surf.}}$. The mean and standard deviation of $\varepsilon_{111,\mathrm{surf.}}$ are plotted in Figure \ref{fig:Fig2}c. The reduction in average $\varepsilon_{111,\mathrm{surf.}}$ with increasing charging time suggests a small average compression at grain boundaries. This is consistent with the accumulation of hydrogen at grain boundaries, leading to a slight compression of near-boundary material. Another interpretation could be the hydrogen-facilitated diffusion of chromium to the grain boundary to replenish any chromium lost during the possible formation of carbides during the annealing process, resulting in a reduction of tensile strain. 

\subsection{Dislocation dynamics}\label{subsec:Dislocation dynamics and simulation}

\begin{figure}[!ht]
    \centering
    \includegraphics[width=0.5\linewidth]{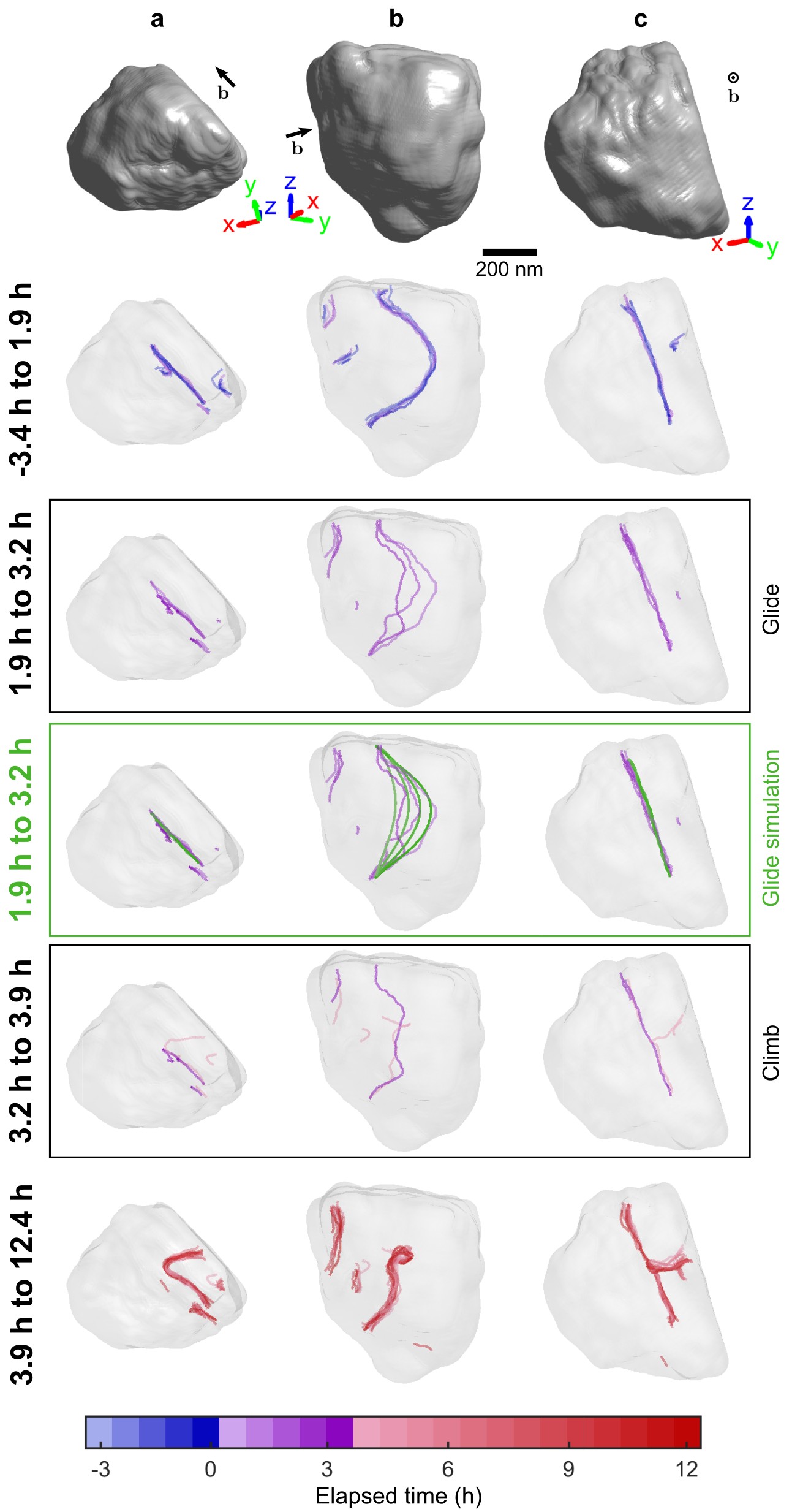}
    \caption{Evolution of the dislocations before and during hydrogen charging. a,b,c) Columns represent a different orthogonal view of the grain and the dislocations. The top row shows the initial morphology of the grain rendered as a grey isosurface. The subsequent rows have translucent renderings of the same grain morphology (largely unchanged, see Figure \ref{fig:Fig2}) along with the dislocations at different states, with glide and climb events indicated. The dislocation dynamics simulation of glide is indicated in green. Dislocations are colored according to the time of observation. The black arrow indicates the Burgers vector. See Video S2 (Supporting Information) for the dislocations at each measurement time.}
    \label{fig:Fig3}
\end{figure}

A phase vortex in a BCDI reconstruction indicates the spatial position of a dislocation core \cite{Clark2015}, enabling the mapping of 3D dislocations (Experimental Section). \textbf{Figure \ref{fig:Fig3}} shows the dislocation evolution during hydrogen charging. Initially, the grain shows two small dislocations close to the surface, and one large bow-shaped dislocation that extends across the grain, hereafter referred to as the ``large dislocation." The large dislocation remains unchanged before hydrogen charging and up to 1.9 h after the start of charging. To determine its Burgers vector, we assume that it is a shear loop and lies in the $\left(\bar{1}11\right)$ plane, as expected for FCC metals (we confirm the angle between the dislocation glide plane normal and $\mathbf{Q}_{111}$ is 68.9°, less than 2° from the angle between theoretical $\langle111\rangle$ directions and within error). Thus, we determine the orientation matrix, $\mathbf{UB}$ (Supporting Information). Glide dislocations in the $\left(\bar{1}11\right)$ plane can have three different Burgers vectors $\mathbf{b} = \frac{a_0}{2}[110]$, $\frac{a_0}{2}[101]$ or $\frac{a_0}{2}[0\bar{1}1]$. However, $\mathbf{Q}_{111} \cdot \mathbf{b} \neq 0$ only for $\frac{a_0}{2}[110]$ and $\frac{a_0}{2}[101]$, meaning that only dislocations with these Burgers vectors will be visible in a $\mathbf{Q}_{111}$ crystal reflection \cite{Hofmann2020}. By comparing the measured strain field of the large dislocation to a 3D elastic model and discrete dislocation dynamics (DDD) simulations, we determine that the large dislocation has Burgers vector $\mathbf{b} = \frac{a_0}{2}[110]$ (Supporting Information, Figures \ref{supp-fig:dislocation model slices} and \ref{supp-fig:dislocation dynamics models}, and Table \ref{supp-table:error}, Supporting Information).

After 1.9 h of hydrogen charging, the large dislocation begins to glide (Figure \ref{fig:Fig3}). This is remarkable, as Figure \ref{fig:Fig2} shows little change in surface strain, and hence externally applied stress that could drive plasticity. It suggests that hydrogen either reduces the Peierls barrier for dislocation glide, or unpins the dislocation from obstacles such as alloying elements \cite{Anderson2017}. Either way, the introduction of hydrogen enables the dislocation to relax, reducing its line length and thus strain energy (Supporting Information and Figure \ref{supp-fig:dislocation elastic energy}, Supporting Information) \cite{Kirchheim2007a}. A DDD simulation (Experimental Section) is setup with the large dislocation in the position observed before hydrogen charging. The ends of the dislocation are pinned at the grain boundary, consistent with our experimental observation, and the system was then allowed to relax under self-stress. The subsequent dislocation evolution, shown in Figure \ref{fig:Fig3}, agrees well with that observed experimentally. This confirms that the dislocation glide observed in experiments is relaxation-driven, suggesting hydrogen-enhanced dislocation mobility and/or depinning. 

Following the glide motion, after 3.2 h of hydrogen charging, one end of the large dislocation becomes unpinned and climbs out of the glide plane. Subsequently, from 3.9 h onward, the dislocation remains largely unchanged upon further charging. The climb force on a mixed dislocation is the sum of an elastic interaction force and an osmotic force \cite{Anderson2017}. Figure \ref{fig:Fig2}c shows that the average strain on the grain surface changes little, indicating that the contribution of external stresses to dislocation climb will be small. Furthermore, the unpinned dislocation segment is $\approx 200$ nm from the nearest observable small dislocation, meaning that any dislocation--dislocation forces will be minor \cite{Birnbaum1994}. Thus, the climb force on the large dislocation will be dominated by the osmotic force, $F_{\mathrm{os}}$. $F_{\mathrm{os}}$, per unit length of a dislocation line is \cite{Anderson2017},

\begin{equation} \label{eq:osmotic}
    \frac{F_{\mathrm{os}}}{L} = -\frac{k_bTb_e}{\upsilon_a}\ln{\frac{c}{c^0}},
\end{equation}

\noindent where $\upsilon_a$ is the atomic volume, $c$ is the vacancy concentration, $c^0$ is equilibrium concentration of vacancies, and $b_e$ is the edge component of the Burgers vector, given by

\begin{equation} \label{eq:b_e}
    b_e = |\mathbf{b}\times\mathbf{\xi}|,
\end{equation}

\noindent where $\mathbf{\xi}$ is the unit dislocation line segment direction. Here we observe climb mobility to be much higher than expected for FCC materials \cite{Abu-Odeh2020}. Climb requires diffusion of self-interstitial atoms or vacancies to the dislocation line. Our results suggest hydrogen-facilitated climb, driven by the formation and migration of vacancies as suggested by the SAV model \cite{Fukai2003,Nagumo2001}. Furthermore, hydrogen has been demonstrated to enhance metal atom diffusion \cite{Fukai2003} and lower the vacancy migration free energy barrier in FCC metals \cite{Du2020}, which would also drive hydrogen-facilitated climb.



\begin{figure}[!ht]
    \centering
    \includegraphics[width=0.8\linewidth]{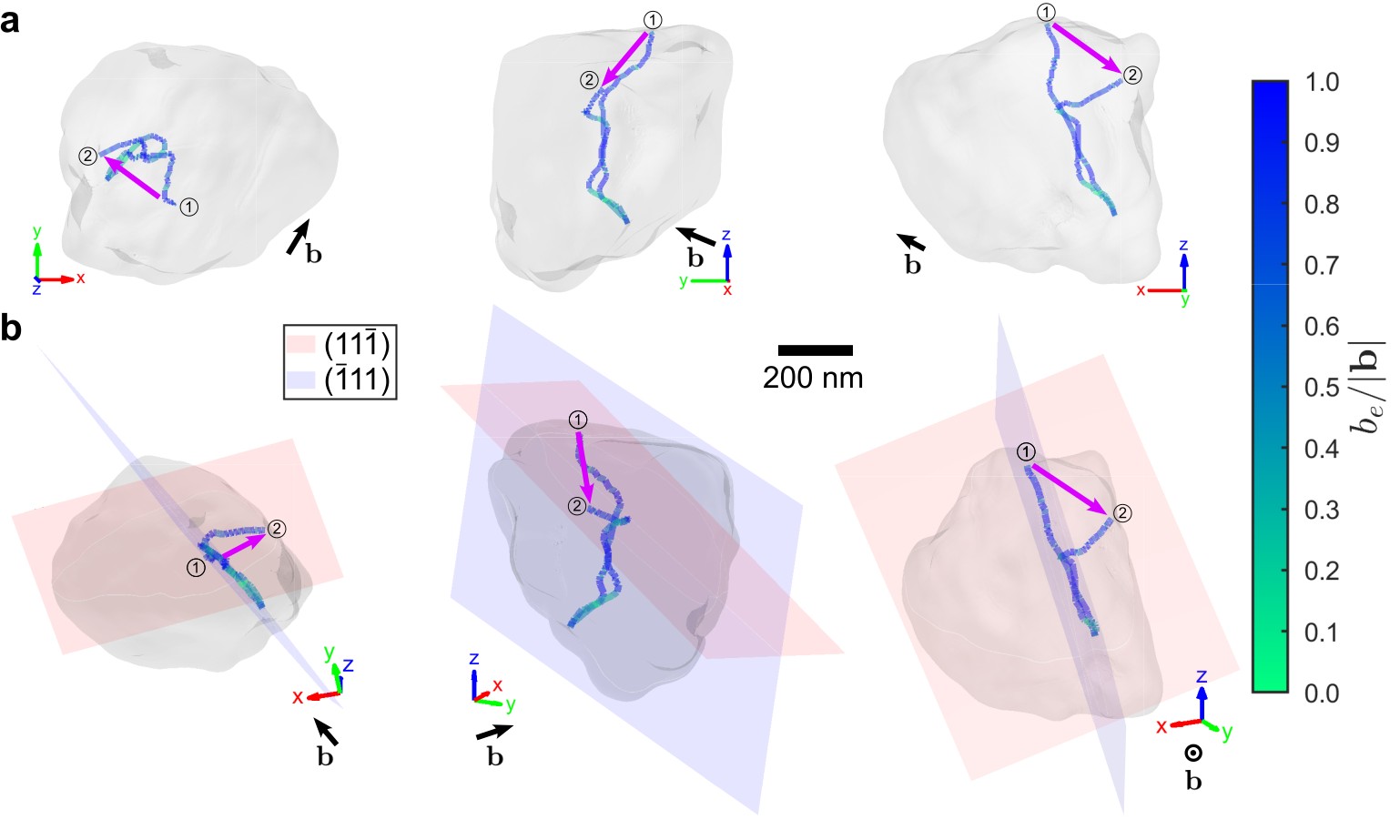}
    \caption{Dislocation unpinning and climb. A translucent isosurface of the grain showing only the large dislocation before (1, at 3.2 h) and after the climb event (2, at 3.9 h). The dislocation is colored between two nodes based on the $b_e$ (Equation \ref{eq:b_e}), where $b_e/|\mathbf{b}| = 1$ represents a pure edge dislocation and $b_e/|\mathbf{b}| = 0$ represents a pure screw dislocation. The black arrow indicates the Burgers vector direction. The magenta arrow, which is nearly perpendicular to the Burgers vector direction, indicates the direction of climb. a) Orthogonal views based on sample coordinates. b) Crystallographic views based on the dislocation loop plane, with the dislocation initially lying on the $\left(\bar{1}11\right)$ plane, and later one end unpins and climbs onto the $\left(11\bar{1}\right)$ plane.}
    \label{fig:Fig4}
\end{figure}

Our observations suggest that once there is a sufficient hydrogen concentration within the grain, there will be enough vacancies and sufficient reduction in vacancy and metal migration energy such that the osmotic force can facilitate climb. The equilibrium concentration of vacancies along the length of the large dislocation depends on the spatial variation of hydrogen concentration. Based on the magnitudes of $e_{111}$ $\left(\approx 10^{-3}\right)$ compared to $\varepsilon_{111}$ $\left(\approx 10^{-4}\right)$, increasing hydrogen concentration dominates homogeneous lattice swelling. Therefore, any variations in hydrogen concentration along the dislocation, and by extension, vacancy concentration, are small by comparison. This suggests that differences in $b_e$, i.e. the degree of edge character along the dislocation line, will determine which parts of the dislocation climb. \textbf{Figure \ref{fig:Fig4}} shows a magnified view of the large dislocation before and after the climb at 3.9 h. Each segment of the dislocation is colored by edge character. The edge character is low at the end of the dislocation that does not climb, making this segment less likely to climb due to a lower $F_{\mathrm{os}}$. Conversely, the segment of the dislocation that climbs also has the greatest edge character. The climb direction (magenta arrow in Figure \ref{fig:Fig4}) is nearly perpendicular (89.4°) to the Burgers vector, as expected.

After climb, the climbing segment lies on the $\left(11\bar{1}\right)$ plane. However, since $\mathbf{b} = \frac{a_0}{2}[110]$ is not one of the Burgers vectors in this plane, this segment is sessile, explaining the lack of further evolution beyond 3.9 h. This sudden dislocation pinning may result from a sufficiently high hydrogen concentration accumulating at the dislocation core \cite{Murakami2010}, as recently observed in ferritic steel using ECCI \cite{Kim2025}. Alternatively, high concentrations of hydrogenated vacancies could lead to dislocation immobilization, as shown in other in situ TEM experiments on initially mobile dislocations in aluminum \cite{Xie2016}.

Dislocation climb is surprising since dislocation cross-slip is more energetically favored. Dislocation cross-slip requires dislocation screw character. Here, the Burgers vector is $\mathbf{b} = \frac{a_0}{2}[110]$, which is not parallel to the dislocation line segment that unpins, thus the segment cannot cross-slip from the $\left(\bar{1}11\right)$ plane to the $\left(11\bar{1}\right)$ plane (Figure \ref{fig:Fig4}). Unfortunately, we do not know what the dislocation looks like between 3.2 and 3.9 h, so we cannot rule out cross-slip based on the absence of screw character alone. However, cross-slip also requires high stacking fault energy, which is known to be reduced by hydrogen in austenitic SS \cite{Hatano2014}. It also requires applied stress, which is minimal based on Figure \ref{fig:Fig2}. The proximity of free surfaces can enable spontaneous cross-slip nucleation in FCC materials \cite{Rao2013}, however, based on the $\approx2\mu $m depth of the embedded grain, this phenomenon is unlikely as image stresses are small at this depth \cite{Yoffe1961}. Thus, we propose that vacancy-assisted climb is most plausible in the hydrogen environment.

\subsection{Dislocation strain field evolution}\label{subsec:Dislocation strain field and modelling}

\begin{figure}[!ht]
    \centering
    \includegraphics[width=0.7\linewidth]{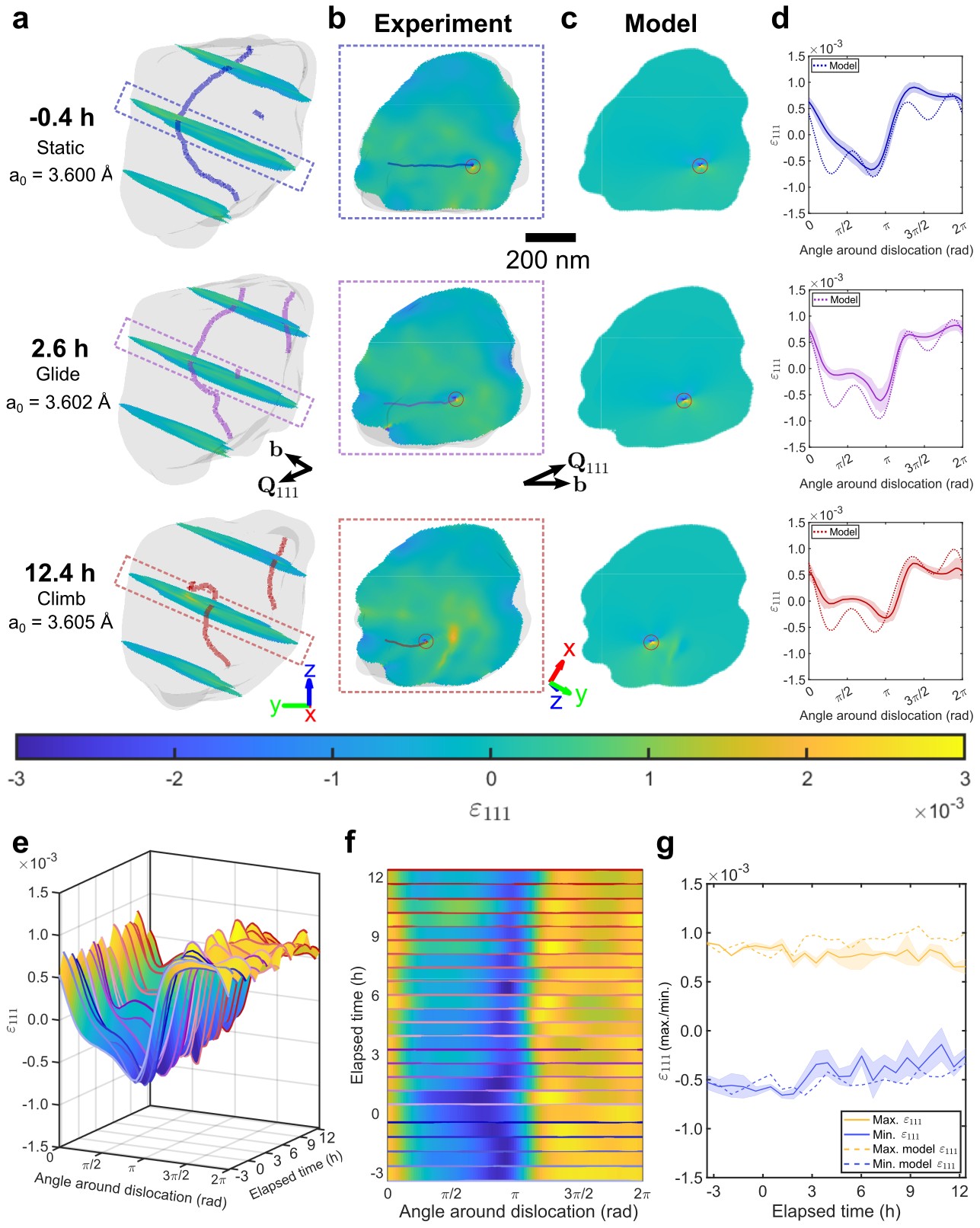}
    \caption{Evolution of the internal strain field surrounding the large dislocation. Three time points are presented as rows in (a--d). a) Translucent morphologies of the grain with dislocations. Slices through $\varepsilon_{111}$ are in-plane to the Burgers vector and the normal to the $\left(\bar{1}11\right)$ plane. The average lattice parameter is listed for each time. b) A $\varepsilon_{111}$ slice capturing a section of the large dislocation with pure edge character. See Video S3 (Supporting Information) for $\varepsilon_{111}$ slices through the entire grain. c) Theoretical elastic model, $\varepsilon_{111,\ \mathrm{model}}$, of the large dislocation devoid of hydrogen (Experimental Section). d) Circular line profiles, drawn at a 30 nm radius from each dislocation core, compared to the model. The shaded region is the standard deviation of the experimental values in the neighboring 26 pixels. Line profiles start at the Burgers vector (horizontal) and run anticlockwise. e) Circular line profiles drawn for each time point of the experiment, plotted as a surface indicating $\varepsilon_{111}$ values. f) Top-down view of (e). g) Maximum and minimum values of the line profiles, averaged over a range of $\pi/4$. The shaded region corresponds to one standard deviation of each value. To better highlight the differences between the model and experimental data at positive charging times, the model's maximum and minimum values were slightly offset to align with the experimental data at negative times.}
    \label{fig:Fig5}
\end{figure}

Based on observations of dislocation pileup evolution upon introduction of hydrogen, it has been previously inferred that hydrogen reduces the interaction between dislocations by changing their stress fields \cite{Birnbaum1994,Ferreira1998}. To quantify the changes in the strain field surrounding the large dislocation, we compare slices though the 3D reconstructed volume where $\mathbf{b}$ and $[\bar{1}11]$ (the normal to the initial glide plane of the large dislocation) are in plane and $\xi$ is normal to the slice  (i.e. where the large dislocation has predominantly edge character). These slices are shown in \textbf{Figure \ref{fig:Fig5}}a, b. A theoretical model, devoid of hydrogen, of the edge dislocation strain field projected along $[111]$ is shown in Figure \ref{fig:Fig5}c (Experimental Section), and compared to the experimental strain field by considering circular line profiles drawn at a 30 nm radius from the dislocation core (Figure \ref{fig:Fig5}d). Agreement between the theoretical model and the experiment is initially good, but worsens as the hydrogen concentration increases.

The evolution of strain along the circular line profiles during charging is shown in Figure \ref{fig:Fig5}e,f. Figure \ref{fig:Fig5}g shows how the maximum and minimum values, averaged over $\pm\pi/4$ around each maximum/minimum, evolve during charging. We observe a reduction of the elastic strain surrounding the dislocation by up to 35\% between the start and end of hydrogen charging. This relative reduction agrees well with the computed reduction of edge dislocation-carbon atom interaction energy in $\alpha$-Fe upon the introduction of hydrogen \cite{Sofronis1995}. 

At the nanoscale, we attribute this elastic strain reduction to the formation of a hydrogen atmosphere in the vicinity of the dislocation. Dislocations are relatively large and strong hydrogen trapping sites ($\approx 27\ \mathrm{kJ\ mol^{-1}}$) \cite{Choo1982a}, so the hydrogen atoms surrounding the dislocations are relatively immobile at room temperature \cite{Myers1992}. Therefore, the evolution of this localized accumulation can be probed using the heterogeneous strain. Figure \ref{fig:Fig5}g also shows that the maximum $\varepsilon_{111}$ deviates more from the elastic hydrogen-free model than the minimum $\varepsilon_{111}$. This suggests that hydrogen prefers to reside in the tensile region of the dislocation \cite{Birnbaum1994}. Alternatively, the reduction in tensile strain could also be due to the hydrogen-facilitated diffusion of solutes, resulting in solute segregation \cite{Kirchheim2007a}, but this is unlikely since we are at room temperature.

Our observations reveal hydrogen elastic shielding (Figure \ref{fig:Fig5}) which has been predicted by simulations \cite{Birnbaum1994}, and indirectly inferred from in situ TEM observations of dislocation pileup evolution following the introduction of hydrogen \cite{Ferreira1998}. These direct BCDI measurements of subtle changes in dislocation strain field are important, since they provide fundamental validation of hydrogen-induced dislocation stress shielding, which is core to the HELP mechanism. 

\subsection{Proposed mechanisms}\label{sec:Proposed mechanisms}

We now consider whether dislocation mobility can be introduced by internal stresses induced by hydrogen concentration gradients caused by cathodic hydrogen charging \cite{Rozenak1983}. As hydrogen diffuses into the SS disk from the top, lattice swelling ensues \cite{Ulmer1993,Fukai2005}, creating a hydrostatic internal stress. There is an in-plane boundary constraint from the uncharged bulk that generates an in-plane stress on the grain, such that the in-plane strain remains zero, as required for compatibility \cite{Hofmann2015}. We have modeled the experimental conditions and predicted the hydrogen concentration as a function of time and depth below the surface of the disk (Supporting Information and Figure \ref{supp-fig:H_concentration_profile}a, Supporting Information). The resulting strain distributions can be compared to the experimental data. In Fig. \ref{supp-fig:Strain_comparison}, Supporting Information, we observe that there is an additional background strain field in the measured $\varepsilon_{111}+e_{111}$ data that is not directly captured by the hydrogen-induced strain gradient combined with the large dislocation strain field model, $\varepsilon_{111,\ \mathrm{model}}+\varepsilon_{111,\ \mathrm{b.c.}}$. We do note that this background remains largely unchanged throughout the charging, suggesting that stresses due to neighboring grains remain largely unchanged during hydrogen uptake.

The hydrogen uptake creates a resolved shear stress that acts on the dislocation. The in-plane stress has been applied in the DDD model, which shows only a slight lift in part of the dislocation toward the unpinning direction (orange arrows in Figure \ref{supp-fig:dislocation dynamics models stress}, Supporting Information). Computation of the resulting Peach--Koehler (PK) force on the large dislocation reveals a significant component in the out-of-plane direction of dislocation climb (purple arrows in Figure \ref{supp-fig:Fig4_bc}, Supporting Information). Interestingly, the PK force also has an in-plane component opposite the direction of glide, which explains why the DDD model does not exhibit gliding behavior in Figure \ref{supp-fig:dislocation dynamics models stress}, Supporting Information. This allows us to conclude that, while hydrogen gradient-induced stresses are significant, they do not appear to be the primary factor controlling dislocation glide. Instead, the hydrogen stress shielding effect \cite{Ferreira1998} reduces the large dislocation's effective Burgers vector, lowering the PK force and creating an energetically favorable condition where the dislocation line straightens, as we observe experimentally.

Our results capture the evolution of a single dislocation undergoing glide relaxation, followed by climb. We propose that this sequence is initiated by the introduction of hydrogen into the lattice, which greatly enhances dislocation mobility via several complementary mechanisms. A hydrogen-induced reduction of lattice friction, or Peierls stress, has been predicted by DFT studies on pure metals using the Peierls--Nabarro model \cite{Kumar2021}. However, in 316 SS, this reduction is small compared to the pinning stress from alloying elements, given that 316 SS's yield stress is more than twice that of pure iron. In addition, hydrogen accumulation has been predicted to lower the dislocation core energy \cite{Yu2020} and reduce the dislocation strain field \cite{Sofronis1995}, as we have directly measured (Figure \ref{fig:Fig5}). This reduces the elastic interactions of the dislocation with the surrounding environment, such as other dislocations and point defects, allowing easier dislocation glide. Furthermore, our observation that hydrogen greatly enhances dislocation climb mobility (Figure \ref{fig:Fig4}) suggests that also during glide, dislocations will be able to ``climb around" obstacles that would otherwise lead to dislocation pinning. Together, these observations suggest that, in the presence of hydrogen, the currently accepted view of glide-dominated dislocation motion \cite{Birnbaum1994,Robertson2015,Ferreira1998} may no longer hold.

\section{Conclusion}

Our findings showcase the unique capability of in situ BCDI to monitor the evolution of nanoscale deformation and defects in a grain exposed to hydrogen under bulk conditions. This advancement is particularly timely as third-generation synchrotrons are being upgraded to fourth-generation facilities, offering orders of magnitude greater coherent flux for BCDI experiments that will enhance spatial and temporal resolution. This could enable exciting possibilities for exploring hydrogen-induced evolution in adjacent grains or extended grains using Bragg ptychography \cite{Godard2011}. In situ BCDI is applicable to most alloy systems and perfectly complements higher-resolution electron microscopy techniques, computational models, and atomistic simulations to shed light on HE degradation pathways. Using the example of 316 SS, our results show the hydrogen-driven evolution of a single dislocation undergoing glide, followed by climb. We propose that glide is initiated by the diffusion of hydrogen into the lattice, where it reduces pinning strength and thus enables the dislocation to glide in-plane, reducing its line length and thus elastic energy. Our strain measurements also show that hydrogen shields the elastic strain field associated with the dislocation. With increasing hydrogen concentration, we observe rapid dislocation climb that is not expected at room temperature. This suggests that hydrogen alters the energy landscape, such that vacancies and interstitials become more readily available at the dislocation line, thus enabling climb motion. Understanding these internal mechanisms is key to designing alloys with enhanced resistance to hydrogen-induced degradation, such as engineering micro- and nano-structures to mitigate dislocation movement or vacancy formation \cite{Li2020c}. These hydrogen-resistant materials are urgently needed to enable the green hydrogen economy to help achieve decarbonization targets.

\section{Experimental Section}

\subsection{Sample preparation}\label{subsec:HPT}
The SS sample was produced using high pressure torsion (HPT) to produce sub-micron grain sizes appropriate for BCDI. A sheet of 316 A4 austenitic stainless steel (composition shown in Table \ref{supp-table:composition}, Supporting Information), 1 mm thick, was obtained from RS Components (stock number 264-7241) and cut into a 5 mm diameter disk. The disk then went through HPT to refine the grain microstructure. HPT was performed in a quasi-constrained set-up \cite{Zhilyaev2008} on a Zwick Roell Z100 Materials Testing machine.  Using 5 mm diameter anvils, a compressive force of 80 kN was applied to the face of the disk and maintained for 30 turns using a rotational speed of 5 ° $\mathrm{s^{-1}}$. The final disk thickness after HPT was 300 $\mu $m. The disk was then annealed in vacuum ($5\times10^{-6}$ mbar) at 700 °C for 1 h, ramping up and down at 4 °C $\mathrm{min}^{-1}$. Following annealing, the disk was sequentially ground with 800, 1200, and 2500 grit SiC paper, followed by polishing in 3 $\mu $m and 1 $\mu $m diamond suspension. 

Powder X-ray diffraction (Figure \ref{supp-fig:powder_XRD}, Supporting Information) showed that after HPT, the sample was predominately composed of deformation-induced martensite. The heat treatment removed macroscale residual stresses and the sample underwent recovery and recrystallization to form austenite \cite{Bartova2022,Naghizadeh2018}. The measured austenite grain was likely formed during heat treatment due to the relatively low misorientation and very low observed strain (relative to the average lattice parameter of the grain) near the surface of the grain. It was noted that carbides could have formed during the annealing, which could attract solutes that have become mobile due to hydrogen. 

\subsection{BCDI measurements}\label{subsec:BCDI}
BCDI relies on inverting far-field, oversampled \cite{Sayre1952}, 3D coherent X-ray diffraction patterns (CXDPs) for a particular $hkl$ reflection from a finite crystalline material. This was done using phase retrieval algorithms \cite{Fienup1982} to obtain the grain morphology and associated phase, $\mathbf{\psi}_{hkl}(\mathbf{r})$. Dislocation line positions can be readily identified from the positions of phase vortex loci \cite{Clark2015}, thus making BCDI a valuable tool to study in situ or in operando defect evolution \cite{Clark2015,Orr2023,Singer2018,Yau2017b,Gorobtsov2023}. The strain fields associated with the dislocations can be obtained from the phase, $\mathbf{\psi}_{hkl}(\mathbf{r})$, since $\mathbf{\psi}_{hkl}(\mathbf{r})$ is a projection of the lattice displacement field, $\mathbf{u}_{hkl}(\mathbf{r})$, onto the scattering vector, $\mathbf{Q}_{hkl}$ \cite{Robinson2009}.

BCDI was performed at beamline 34-ID-C at the Advanced Photon Source (APS) at Argonne National Laboratory, USA. An in situ confocal microscope was used to position the X-ray beam within the disk. The $111$ Bragg peak ($2\theta = 34.7$°) was measured for multiple grains to screen for the best candidate grain. Although each candidate's position perpendicular to the X-ray beam could be easily aligned, the position of the candidate along the beam needs to be determined to ensure the grain is on the axis of rotation. An approach presented by Shabalin et al. \cite{Shabalin2023} was used to accomplish this.

For all BCDI measurements, the grain was illuminated using a 10 keV ($\lambda$ = 0.124 nm) coherent X-ray beam, with a bandwidth of $\Delta\lambda/\lambda \approx 10^{-4}$ from a Si$(111)$ monochromator. The X-ray beam was focused to a size of 810 nm × 860 nm (h × v, full width at half-maximum) using Kirkpatrick--Baez (KB) mirrors. Beam defining slits were used to select the coherent portion of the beam at the entrance to the KB mirrors. CXDPs were collected on a 256 × 256 pixel module of a  512 × 512 pixel Timepix area detector (Amsterdam Scientific Instruments) with a GaAs sensor and pixel size of $55 \mathrm{\ \mu m} \times 55 \mathrm{\ \mu m}$ positioned at 1.4 m from the sample to ensure oversampling. The peak of the CXDP was positioned at the center of the detector module before data collection. CXDPs were recorded by rotating the crystal through an angular range of 0.5° about the peak and recording an image every 0.005° with 0.5 s exposure time and 10 accumulations at each angle.

\subsection{Hydrogen charging}\label{subsec:Hydrogen charging}
The disk was attached to the electrochemical flow cell using chemically resistant insulating tape, with the bottom of the disk in contact with a Pt wire, thereby forming the working electrode (WE). The counter electrode (CE) was another Pt wire coiled into the well, or moat, of the electrochemical cell. The cell was then sealed using an O-ring and a thin kapton film. The flow cell was fixed to the sample stage using a Thorlabs 1X1 kinematic mount and connected to a Cole-Parmer Masterflex peristaltic pump and a SP-300 Biologic potentiostat. See Supporting Information and Figure \ref{supp-fig:flowcell}, Supporting Information for further details about the electrochemical flow cell.

A 1 L hydrogen charging solution was prepared using 4 g of NaOH (0.1 mol $\mathrm{L}^{-1}$) and 5 g thiourea (0.07 mol $\mathrm{L}^{-1}$) dissolved in deionized water, yielding a pH of 13. After an appropriate grain was found, the hydrogen charging solution was continuously pumped into the cell at 5 mL $\mathrm{min}^{-1}$. Repeated $111$ CXDPs from the grain were measured in solution, without hydrogen charging, over three hours to verify that there were no X-ray effects on the sample. Furthermore, any X-ray induced changes to the strain field are negligible (Supporting Information). Next, hydrogen charging was performed using chronopotentiometry with a two electrode setup, keeping the current fixed at 0.2 mA (0.5 mA $\mathrm{cm^{-2}}$). During the charging, two repeated BCDI measurements for the $111$ reflection were measured approximately every 40 min, with each scan requiring approximately 10 min to complete. Bubbles were noticed to slowly form during charging.

\subsection{Phase retrieval}\label{subsec:Phase retrieval}
Before phase retrieval, the repeated CXDPs were corrected for dead-time, darkfield, and whitefield before cross-correlation alignment and summation. The minimum data threshold was 2. The resulting CXDP was binned by a factor of two along each direction of the detector plane, resulting in a size of $128 \times 128 \times 101$ voxels.

The reconstructions were processed using a MATLAB phase retrieval package \cite{Clark2015}. The reconstructions for each time point were seeded using a reconstruction from eight repeated CXDPs (combined following the same procedure) of the grain before the charging solution was pumped into the flow cell. This seed was reconstructed starting with a random guess. A guided phasing approach \cite{Chen2007} with 100 individuals and four generations was used with a geometric average breeding mode and a low to high-resolution scheme \cite{McCallum1989}. For each generation and population, a block of 20 error reduction (ER) and 180 hybrid input-output (HIO) iterations, with $\beta = 0.9$, was repeated three times. This was followed by 20 ER iterations to return the final object. The shrinkwrap algorithm \cite{Marchesini2003} with a threshold of 0.08 was used to update the real-space support every iteration. The best reconstruction was determined using a sharpness criterion, appropriate for crystals containing defects \cite{Ulvestad2017b}. The seed reconstruction was then used as the initial guess for the reconstructions at all time points, which followed the same guided phase retrieval procedure as the seed.

\subsection{BCDI strain calculations}\label{subsec:BCDI strain}
The residual, or heterogeneous strain relative to the average lattice, $\varepsilon_{hkl}$, is calculated by,

\begin{equation} \label{eq:het_strain}
   \varepsilon_{hkl}\mathbf{(r)} = \frac{\partial \mathbf{u}_{hkl}(\mathbf{r})}{\partial x_{hkl}} = \nabla\mathbf{\psi}_{hkl}(\mathbf{r})\cdot\frac{\mathbf{Q}_{hkl}}{|\mathbf{Q}_{hkl}|^2}
\end{equation}

\noindent This differs from the homogeneous strain, $e_{hkl}$, associated with changes of the average lattice parameter at time $t$, $a_{0,\ t}$,

\begin{equation} \label{eq:homo_strain}
    e_{hkl} = \frac{a_{0,\ t}-a_{0\mathrm{,\ ref.}}}{a_{0\mathrm{,\ ref.}}}
\end{equation}

\noindent where $a_{0\mathrm{,\ ref.}}$ is the reference lattice parameter, determined at --3.4 h. $a_{0}$ was calculated using Bragg's law and the $2\theta$ angle determined from the angular position of the diffractometer.

\subsection{Hydrogen concentration}\label{subsec:Hydrogen concentration}
The volumetric strain, $\varepsilon_{\mathrm{vol}}$, associated with the swelling of the grain due to hydrogen uptake can be written as \cite{Hofmann2015},

\begin{equation} \label{eq:eq_vol}
    \varepsilon_{\mathrm{vol}} = \sum_A C^{(A)}\Omega_R^{(A)},
\end{equation}

\noindent where $C^{(A)}$ is the defect concentration for defect type $A$, and $\Omega_R^{(A)}$ is the relaxation volume for the specific defect type. The FCC metal lattice has two interstitial sites available for accommodating hydrogen atoms: octahedral (O) and tetrahedral (T) sites. In transition metals with FCC lattice, dissolved hydrogen atoms preferentially occupy the O site with larger free space than the T site \cite{Fukai2005}. $\Omega_R = 0.200 \pm 0.005$ was used for interstitial hydrogen in austenitic stainless steels \cite{Ulmer1993}. For the calculation, this value was assumed to be similar for FCC 316 SS, and that all the interstitial hydrogen atoms reside in the O sites.


The volumetric strain relative to the initial volume at --3.4 h was calculated using the homogeneous strain, $e_{hkl}$ (Equation \ref{eq:homo_strain}),

\begin{equation} \label{eq:eq_vol_e}
    \varepsilon_{\mathrm{vol}} = (1+e_{hkl})^3-1
\end{equation}

\noindent Thus, the concentration of hydrogen was determined as, 

\begin{equation} \label{eq:H_concentration}
    C = \frac{(1+e_{hkl})^3-1}{\Omega_R}
\end{equation}

\noindent The hydrogen concentrations presented in Figure \ref{fig:Fig1}b are reasonable when compared to values obtained from charging hydrogen into 316L SS \cite{Brass2006}.

\subsection{Pearson correlation coefficient}\label{supp-sec:pearson}
The Pearson correlation coefficient, $r$, between two images was computed using Equation (\ref{eq:XC}):

\begin{equation}\label{eq:XC}
     r(x,y) = \frac{\sum\limits_{n}(x_n-\bar{x})(y_n-\bar{y})}{\sqrt{\sum\limits_{n}(x_n-\bar{x})^2}\sqrt{\sum\limits_{n}(y_n-\bar{y})^2}}
\end{equation}

\noindent where $x$ and $y$ are the Bragg peaks being compared, $x_n$ and $y_n$ are the values for a single voxel, and $\bar{x}$ and $\bar{y}$ are the means of each array.

\subsection{Dislocation position identification}\label{subsec:Dislocation analysis}
The dislocation line is identified as the spatial positions of the locus of the phase vortices, referred to as nodes. Each dislocation was composed of many nodes, joined by edges based on the MATLAB graph object. The nodes were determined automatically by integrating the derivatives of the complex exponential of the phase $(e^{i\psi_{hkl}(\mathbf{r})})$ \cite{Yang2022c} and then selecting the maximum value as the dislocation node position if the value exceeded $0.75\pi$. The dislocation lines were found not to overlap perfectly in the reconstructions, which were attributed to noise and the limited spatial resolution. To increase overlap, the reconstructions were subpixel-shifted \cite{Guizar-Sicairos2008}, on the order of a few pixels, or up to $\approx 30$ nm. The dislocation lines terminated on the surface of the grain.

\subsection{Dislocation dynamics modeling}\label{subsec:Dislocation dynamics}

\subsubsection{Overview}
DDD was used to simulate the motion of the dislocation structure and compare to the experimental result.
The DDD simulations here are nodal, based on discrete straight segments \cite{Arsenlis:MSMSE:2007,Cai:MSEA:2004} presented by Yu et al. \cite{Yu:JotMaPoS:2019}.
Dislocation motion is calculated by evaluating the velocity $\textbf{V}_k$ of node $k$ at position $\textbf{X}_k$ through a mobility law (shown in the following section), which describes mobility in various modes like glide, climb, and cross-slip.




\subsubsection{FCC Mobility Law}
The nodal force, $\textbf{F}^k$, of node $k$ at position $\textbf{X}^k$ is evaluated at each time increment for every node, and the dislocation structure is updated using the nodal velocity $\textbf{V}^k$, as computed through an FCC mobility law.
The nodal force, $\textbf{F}$, has contributions from the segment-segment interaction force, obtained from the non-singular dislocation stress field, the dislocation core force, and the forces due to the corrective (image) stress field.
The total nodal force can therefore be represented as

\begin{align}
    \boldsymbol{F}^k &= \sum_l \sum_{i,j} \tilde{\boldsymbol{f}}^{kl}_{ij}(\boldsymbol{X}^k)+\sum_l \boldsymbol{f}^{kl}_c(\boldsymbol{X}^k)+\sum_l \boldsymbol{\hat{f}}^{kl}(\boldsymbol{X}^k) \\
    &= \boldsymbol{\tilde{F}}^k+\boldsymbol{F}^k_c+\boldsymbol{\hat{F}}^k
    \label{eq:nodalforces}
\end{align}

\noindent where $\tilde{\boldsymbol{f}}^{kl}_{ij}(\boldsymbol{X}^k)$ is the interaction force at node $k$, due to segment $i \rightarrow j$ integrated along segment $k\rightarrow l$.
This is summed over all segments $i \rightarrow j$ within the domain, including the self force due to segment $k \rightarrow l$.
Finally this is then summed over all nodes $l$ which are connected to node $k$ to give the interaction force on node $k$, $\boldsymbol{\tilde{F}}^k$. 
The quantity $\boldsymbol{F}^k_c$ is the dislocation core force and $\boldsymbol{\hat{F}}^k$ is the corrective elastic force evaluated with the finite element method using the superposition principle to account for the finite boundary.

For each segment $kl$ with $L^{kl}$, a drag tensor $\boldsymbol{B}^{kl}$ is determined according to the segment character.
The nodal velocity $\boldsymbol{V}^k$ at node $k$ is then calculated as

\begin{equation}
    \Big[\frac{1}{2} \sum_l L^{kl} \boldsymbol{B}^{kl} \Big]^{-1} \boldsymbol{F}^k = \boldsymbol{V}^k
\end{equation}

\noindent where the sum is over all nodes $l$ connected to node $k$, and $\boldsymbol{F}^k$ is the nodal force determined by Equation (\ref{eq:nodalforces}).

All dislocation segments are constrained to the $\{111\}$ slip planes in FCC, and their respective drag tensor can be expressed as \cite{Bulatov:NA:2006}

\begin{equation}
    \boldsymbol{B}^{kl}(\boldsymbol{l}^{kl}) = B_g(\boldsymbol{m}^{kl}\otimes\boldsymbol{m}^{kl})+B_c(\boldsymbol{n}^{kl}\otimes\boldsymbol{n}^{kl})+B_l(\boldsymbol{l}^{kl}\otimes\boldsymbol{l}^{kl})
    \label{eq:dragtensorequation}
\end{equation}

\noindent where $B_g, B_c,$ and $B_l$ are the drag coefficients for glide, climb, and motion along the line direction, respectively.
The unit vectors are the line direction $\boldsymbol{l}^{kl}$, the slip plane normal $\boldsymbol{n}^{kl}$, and glide direction $\boldsymbol{m}^{kl}$.\\

\subsubsection{Incorporating finite boundary conditions}
To evaluate the corrective force term $\boldsymbol{\hat{F}}^k$ in Equation (\ref{eq:nodalforces}), the superposition principle is adopted. 
As described elsewhere \cite{Giessen:MSMSE:1995,Weygand:MSMSE:2002}, the total stress strain and displacement fields are expressed as

\begin{align}
    \boldsymbol{\sigma} &= \boldsymbol{\hat{\sigma}}+\boldsymbol{\tilde{\sigma}} \\
    \boldsymbol{\varepsilon} &= \boldsymbol{\hat{\varepsilon}}+\boldsymbol{\tilde{\varepsilon}} \\
    \boldsymbol{u} &= \boldsymbol{\hat{u}}+\boldsymbol{\tilde{u}}
\end{align}
\noindent respectively. 
The infinite-body fields are denoted as $\tilde{()}$ and the finite-element correction fields as $\hat{()}$.
According to the following procedure \cite{El-Awady:JotMaPoS:2008}, the image stress field may be evaluated.
First, the elastic stress field due to dislocations in an infinite body, $\tilde{\sigma}$, is obtained.
The tractions $\boldsymbol{\tilde{T}} = \boldsymbol{\tilde{\sigma}}\cdot n$ on the traction boundaries due to this stress are then calculated, and subtracted from the desired boundary conditions $\boldsymbol{T}$.
These modified boundary values, $\boldsymbol{\hat{T}} = \boldsymbol{T}-\boldsymbol{\tilde{T}}$, in addition to the displacement conditions $\boldsymbol{U}$ on the displacement boundaries are used in an elastic finite element simulation to determine the corrective fields.
Finally, the corrective stress field, $\boldsymbol{\hat{\sigma}}$, is used to evaluate the corrective nodal force, $\boldsymbol{\hat{f}}$.\\

\subsubsection{Calculation Details}
The initial dislocation structure used in the simulation is identical to the initial dislocation structure measured by experiment before the start of hydrogen charging.
The following parameters were used to represent $\gamma$-Fe: the lattice parameter, $a_0=3.602$~\r{A}, the shear modulus, $G = 77\ $ GPa, Poisson's ratio, $\nu = 0.28$,  
The dislocation structure was enclosed within a 0.6 $\mathrm{\mu}$m $\times$ 0.6 $\mathrm{\mu}$m $\times$ 0.6 $\mathrm{\mu}$m domain representing the experimentally observed grain.
The finite element (FE) mesh was $20\times20\times20$.
The drag coefficients for edge and screw dislocations were both 1.0, essentially making the mobility of each type of dislocation equivalent.
In contrast, the drag coefficient for dislocation segments attempting to move in the climb direction was $10^8$, essentially confining the segments to the glide plane.
The two end points were fixed for the entire simulation, while the internal nodes were free in all directions (though as discussed, they were restricted in the climb direction).
No external mechanical load was applied to the domain.

\subsection{Dislocation strain field modeling}\label{subsec:modelling}
It was assumed that there were no externally applied stresses. The $\varepsilon_{111,\ \mathrm{model}}$ for the large dislocation was created using the dislocation node positions determined from BCDI, the Burgers vector (using the local lattice parameter, $a_{0,\ t}$), and $\nu = 0.28$. Each dislocation node was joined by an edge, which was then connected to another point to form a dislocation triangle. The displacement field for this triangular dislocation loop was determined using the solution developed by Barnett \cite{Barnett1985, Barnett2007}, which was then numerically differentiated to determine the lattice strain field, $\varepsilon_{\mathrm{model}}$,

\begin{equation} \label{eq:edge_dislo_tensor}
    \varepsilon_{\mathrm{model}} = 
    \begin{bmatrix} 
        \varepsilon_{xx,\ \mathrm{model}} & \varepsilon_{xy,\ \mathrm{model}} & \varepsilon_{xz,\ \mathrm{model}}\\
        \varepsilon_{yx,\ \mathrm{model}} & \varepsilon_{yy,\ \mathrm{model}} & \varepsilon_{yz,\ \mathrm{model}}\\
        \varepsilon_{zx,\ \mathrm{model}} & \varepsilon_{zy,\ \mathrm{model}} & \varepsilon_{zz,\ \mathrm{model}} 
    \end{bmatrix}
\end{equation}

\noindent $\varepsilon_{\mathrm{model}}$ was projected to the $[111]$ direction, 

\begin{equation} \label{eq:edge_dislo_projection}
    \varepsilon_{111,\ \mathrm{model}} = \begin{bmatrix}
        a & b & c 
    \end{bmatrix}\begin{bmatrix} 
        \varepsilon_{xx,\ \mathrm{model}} & \varepsilon_{xy,\ \mathrm{model}} & \varepsilon_{xz,\ \mathrm{model}}\\
        \varepsilon_{yx,\ \mathrm{model}} & \varepsilon_{yy,\ \mathrm{model}} & \varepsilon_{yz,\ \mathrm{model}}\\
        \varepsilon_{zx,\ \mathrm{model}} & \varepsilon_{zy,\ \mathrm{model}} & \varepsilon_{zz,\ \mathrm{model}} 
    \end{bmatrix}\begin{bmatrix}
        a \\ b \\ c 
    \end{bmatrix}
\end{equation}

\noindent where $a$, $b$, $c$, transformed the Cartesian coordinate system for the theoretical edge dislocation, to the experimental sample space. These were determined by,

\begin{equation} \label{eq:eigenvalues}
    \begin{bmatrix}
        a \\ b \\ c 
    \end{bmatrix} = \begin{bmatrix}
\mathbf{x^\prime}&\mathbf{y^\prime}&\mathbf{z^\prime}
    \end{bmatrix}^{-1}\begin{bmatrix}
\mathbf{\hat{Q}_{hkl,\ sam}}
    \end{bmatrix}
\end{equation}

\noindent where $\mathbf{\hat{x}^\prime}$ is $\mathbf{\hat{b}_{sam}}$, $\mathbf{\hat{y}^\prime}$, is the dislocation loop plane normal vector (normalized) at --3.4 h, and $\mathbf{\hat{z}^\prime} = \mathbf{\hat{x}^\prime}\times\mathbf{\hat{y}^\prime}$. All vectors in Equation (\ref{eq:eigenvalues}) are column vectors. The subscript, $\mathbf{sam}$, corresponds to the the sample coordinates relative to the standard unit vectors $\mathbf{\hat{x}}$, $\mathbf{\hat{y}}$, and $\mathbf{\hat{z}}$ axes of a 3D Cartesian coordinate system.


\medskip

\medskip
\textbf{Acknowledgements} \par 
F.H. thanks Thomas D. Swinburne and Max Boleininger for insightful comments. D.Y., G.H., and F.H. acknowledge funding from the European Research Council under the European Union's Horizon 2020 research and innovation programme (grant agreement No 714697). K.S. acknowledges funding from the General Sir John Monash Foundation. A.M. acknowledges funding from EDF (Électricité de France). Work at Brookhaven National Laboratory was supported by the U.S. Department of Energy, Office of Science, Office of Basic Energy Sciences, under Contract No. DESC0012704. This research used the Materials Synthesis and Characterization Facility of the Center for Functional Nanomaterials (CFN), which is a U.S. Department of Energy Office of Science User Facility, at Brookhaven National Laboratory under Contract No. DE-SC0012704. Experiments were performed at the Advanced Photon Source, a US Department of Energy (DOE) Office of Science User Facility operated for the DOE Office of Science by Argonne National Laboratory under Contract No. DE-AC02-06CH11357. Work performed at UCL was supported by EPSRC. The authors acknowledge the use of the Advanced Research Computing (ARC) facility at the University of Oxford \cite{Richards2015}. 


\medskip
\textbf{Conflict of Interest}\par
There are no competing interests to declare.

\medskip
\textbf{Author Contributions} \par
D.Y. and F.H. conceptualized the study; D.Y. and M.S. curated the data; D.Y., M.S., E.T., I.K.R., and F.H. performed formal analysis; E.T., I.K.R., and F.H. acquired funding; D.Y., G.H., K.S., R.J.H., W.C., and D.N. performed investigation; D.Y., B.D., K.S., R.J.H., and W.C. were associated with methodology; F.H. was associated with project administration; D.Y., G.H., K.S., A.M, and F.H. acquired resources; D.Y., M.S., E.T., and F.H. were associated with the software; E.T., I.K.R., and F.H. supervised the study; D.Y. and M.S. performed validation; D.Y. and M.S. visualized the study; D.Y., M.S., and F.H. wrote the original draft; All authors reviewed and edited the final manuscript.

\medskip
\textbf{Data Availability Statement}\par
The data that support the findings of this study are openly available in Zenodo \cite{zenodo_ref}.

%
\bibliographystyle{MSP}
\bibliography{library,seif-bib,dataset}

\end{document}